\documentclass{article}
\usepackage{graphicx} % Required for inserting images
\usepackage{amsfonts,amsmath,amssymb,amsthm}
\usepackage{graphicx}
\graphicspath{{images/}}
\usepackage{makecell}
\usepackage[papersize={8.5in,11in}]{geometry}
\usepackage{authblk}
\usepackage{natbib}
\setcitestyle{aysep={}}
\usepackage{xcolor}
\usepackage[colorlinks=true,linkcolor=blue,citecolor=blue]{hyperref}
\def\app#1#2{%
  \mathrel{%
    \setbox0=\hbox{$#1\sim$}%
    \setbox2=\hbox{%
      \rlap{\hbox{$#1\propto$}}%
      \lower1.1\ht0\box0%
    }%
    \raise0.25\ht2\box2%
  }%
}
\def\approxprop{\mathpalette\app\relax}

\title{The Diffusion Limit of Photoevaporation in Primordial Planetary Atmospheres}

\author[1]{Darius Modirrousta-Galian}
\author[1]{Jun Korenaga}

\affil[1]{Yale University, Department of Earth and Planetary Sciences, 210 Whitney Avenue, New Haven, CT 06511, USA; darius.modirrousta-galian@yale.edu}

\date{}
\begin{document}

\maketitle

\begin{center}
    Accepted: ApJ
\end{center}

\section*{\hfil Abstract \hfil}

Photoevaporation is thought to play an important role in the early planetary evolution. In this study, we investigate the diffusion limit of X-ray and ultraviolet induced photoevaporation in primordial atmospheres. We find that compositional fractionation resulting from mass loss is more significant than currently recognized because it is controlled by the conditions at the top of the atmosphere, where particle collisions are less frequent. Such fractionation at the top of the atmosphere develops a compositional gradient that extends downward. Mass outflow eventually reaches a steady state in which hydrogen loss is diffusion limited. We derive new analytic expressions for the diffusion-limited mass loss rate and the crossover mass.

\begin{center}
\textbf{Keywords:} Atmospheric dynamics -- Atmospheric structure --- Aeronomy
\end{center}

\section{Introduction}

Stars are most luminous in their high-energy bands in their first few hundred millions years after formation \citep{Penz2008(1),Penz2008(2),Sanz-Forcada2011}. Newly formed planets with primordial atmospheres efficiently absorb X-ray and ultraviolet (XUV) photons, triggering photoevaporation and the gradual loss of their hydrogen reservoirs. Observations suggest that atmospheric evaporation is prevalent, as evidenced by detections of leaking hydrogen \citep{Joshi2019} and Xenon isotopic ratios on Earth \citep{Porcelli2003b}, and by exoplanet population trends such as the bimodal radial distribution and sub-Jovian desert \citep{Fulton2017}. The physics of XUV-induced mass loss is however, less clear, with two major models being suggested: inviscid hydrodynamic outflow \citep{Tian2005} and diffusion-limited escape \citep{Zahnle2019}. The first model builds on Parker wind theory \citep{Parker1958} and applies it to planetary atmospheres, suggesting that mass loss occurs through free advection and rapidly erodes hydrogen-rich atmospheres \citep{Kubyshkina2018(1),Caldiroli2021}. In contrast, diffusion-limited escape posits that mass loss occurs preferentially for lighter species such as hydrogen. Interactions between the fast-moving hydrogen and the slow-moving heavier species result in an overall decrease in the hydrogen outflow rate, with mass loss being limited by momentum diffusion between different species \citep{Zahnle2019}.

When a hydrogen-rich atmosphere is exposed to X-ray and ultraviolet irradiation, a steep conducting temperature inversion develops in the thermosphere \citep{Gross1972}. Temperatures become sufficiently high for gases to become gravitationally unbound \citep{Opik1963} and be lost through hydrodynamic winds \citep{Sekiya1980,Sekiya1981}. The question is then whether interactions between different species are indeed significant and mass loss becomes diffusion limited \citep{Hunten1987,Zahnle1990,Zahnle2019} or whether they are small and all species are lost equally through free advection \citep{Kubyshkina2018(1),Caldiroli2021}. If fractionation occurs, the upper regions of the atmosphere become preferentially enriched, imposing a diffusive flux that replenishes lost hydrogen in the enriched background gas. In other words, preferential hydrogen loss requires preferential restocking, which cannot occur through bulk advection alone. Mass transport in the upper atmosphere must therefore involve diffusion, lowering mass loss rates by several orders of magnitude.

In the following, we evaluate the effectiveness of fractionation and the activation of diffusion-limited mass loss in planets with primordial atmospheres. The framework presented in this paper is derived from first principles and it applies to all planets with ideal gas atmospheres, though it is most relevant to those with hydrogen-rich primordial envelopes. We begin with reviewing how the diffusion-limited mass loss rate is defined (section~\ref{sec:diffusion_limited}) and derive a new model for the crossover mass (section~\ref{sec:crossover_mass}). In section~\ref{sec:model_comparison}, we apply these models to the upper atmosphere of Earth. The upper atmosphere of Earth, being hydrogen-rich, is an appropriate analog for exoplanets with hydrogen-rich atmospheres, for which compositional data is limited and uncertain. In section~\ref{sec:steady_state}, we demonstrate that compositional fractionation almost always occurs as a result of X-ray and ultraviolet induced photoevaporation. Advection is shown to be more efficient than diffusion, so a hydrogen depleted layer forms at the top of the atmosphere, which grows rapidly until reaching the bottom of the atmosphere. After equilibrium is reached, mass loss becomes diffusion limited. In section~\ref{sec:simulations}, we present and discuss representative simulations of an exoplanet undergoing photoevaporation and its dependence on the chosen mass loss model. We conclude with an overview of how our findings compare to other approaches in the literature.

\section{Model}

\subsection{Defining the diffusion-limited flux}
\label{sec:diffusion_limited}

Diffusion-limited flux is the maximum rate at which a species can traverse through an atmosphere when advection and diffusion are considered concurrently. Despite its name, diffusion-limited mass loss does not imply a lack of advection. Instead, it describes how momentum diffusion between species acts against bulk flow in an advecting binary gas mixture. At the microscopic scale, the fast moving light species collide with slower moving heavy species, resulting in the light species decelerating and the heavy species accelerating. At the macroscopic scale, it appears as if the heavy species exert drag on the lighter species while the lighter species pull on the heavy species. The diffusion-limited mass loss model extends the Euler equations by incorporating interactions between different species \citep[][p.{\,}107]{Chapman1970}. In the traditional approach of \citet{Hunten1973}, the diffusion-limited flux is estimated by evaluating the relative average velocity of a binary gas mixture in a gravitational field \citep{Chapman1970},\begin{equation}
    \vec{v}_{1}-\vec{v}_{2} = -D\left[\frac{n^{2}}{n_{1}n_{2}}\nabla\left(\frac{n_{1}}{n}\right)+\frac{\mu_{2}-\mu_{1}}{\bar{\mu}}\nabla\left(\ln{P}\right) + \alpha_{T}\nabla\left(\ln{T}\right) - \frac{\mu_{1}\mu_{2}}{\bar{\mu} k_{\rm B}T}\left(\vec{a}_{1}-\vec{a}_{2}\right) \right],
\label{eq:chapman}
\end{equation}
where subscripts 1 and 2 are for the light and heavy constituents, respectively, $D$ is the diffusion coefficient, $n$ is the particle number density, $\bar{\mu}$ is the mean molecular mass, $P$ is the pressure, $\alpha_{T}$ is the thermal diffusion factor \citep[not to be confused with the coefficient of thermal diffusivity;][]{Leuenberger2002}, $T$ is the temperature, $k_{\rm B}$ is the Boltzmann constant, and $a$ is the acceleration acting on the particles from external forces. The first term is the concentration gradient, the second is the mass gradient, the third is the temperature gradient, and the fourth is the external force gradient. Equation~\ref{eq:chapman} is evaluated for Earth at the homopause, which is very close to the mesopause, where there is a temperature inversion and $\nabla\left(\ln{T}\right){=}0$. The homopause is the level below which an atmosphere is well-mixed whereas the mesopause is the boundary between the mesosphere and the thermosphere; the homopause and the mesopause are both located at approximately $85~{\rm km}$, though the homopause is known to vary in altitude from $80{-}120~{\rm km}$ \citep{Salinas2016,Liu2021,Swenson2021}.

The planetary magnetic field is small, so the external acceleration for both particles is set by gravity, which is $g{=}GM_{\rm p}/r^{2}$ ($G$ is the gravitational constant, $M_{\rm p}$ is the planetary mass, and $r$ is the radial distance) and therefore cancels out. In its current form, Equation~\ref{eq:chapman} only includes molecular diffusion, which involves the random movement of individual molecules from areas of high concentration to low concentration. In contrast, eddy diffusion involves mixing by turbulence and acts on the compositional gradient term \citep{Catling2017}, thus it has to be included in its numerator. Introducing the mole fraction $\chi$ so that $n_{1}{=}\chi n$ and $n_{2}{=}\left(1{-}\chi\right)n$, and considering only the radial profile, we thus have
\begin{equation}
    v_{1}-v_{2} = -D\left[\frac{1+\frac{K_{\rm zz}}{D}}{\chi\left(1-\chi\right)}\frac{{\rm d}\chi}{{\rm d}r}+\frac{\mu_{2}-\mu_{1}}{\bar{\mu}}\frac{1}{P}\frac{{\rm d}P}{{\rm d}r} \right].
\label{eq:1}
\end{equation}
According to \citet{Hunten1973}, the compositional gradient term is negligible. Moreover, gas is assumed to be in hydrostatic equilibrium at the homopause, so $1/P\left({\rm d}P/{\rm d}r\right)$ is equal to ${-}1/H$, where $H$ is the scale height defined as $H{=}k_{\rm B}T/\left(g \bar{\mu}\right)$,
\begin{equation}
    v_{1}-v_{2} = \frac{Dg}{k_{\rm B}T}\left(\mu_{2}-\mu_{1}\right).
\label{eq:2}
\end{equation}
In the limit when $\mu_{1}{\ll}\mu_{2}$, Graham's law of diffusion suggests that $v_{1}{\gg}v_{2}$, and $v_{1}{-}v_{2}{\approx}v_{1}$. Instead of employing Chapman-Enskog theory \citep{Chapman1970} for calculating the molecular diffusion coefficient, \citet{Hunten1973} substitutes $D$ with the binary diffusion coefficient, $b_{\rm ij}{=}nD$, which is determined experimentally \citep{Marrero1972} and takes the form $b_{\rm ij}{=}AT^{B}$, with $A$ and $B$ being constants. Multiplying both sides of Equation~\ref{eq:2} by the number density of the light species, $n_{1}$, yields the diffusion-limited flux,
\begin{equation}
    \phi_{1} = n_{1}v_{1} = \frac{\chi b_{\rm ij}g}{k_{\rm B}T}\left(\mu_{2}-\mu_{1}\right),
\label{eq:phi_Ht}
\end{equation}
where $\phi_{1}$ is the escaping particle flux. Mass conservation ensures equal mass flow through all atmospheric spherical shells, thus the global mass loss rate is set by mass flow at the homopause (subscript H),
\begin{equation}
    \dot{M}_{1} = \chi_{\rm H} \frac{4 \pi G M_{\rm p} \mu_{1} b_{\rm ij,H}}{k_{\rm B}T_{\rm H}}\left(\mu_{2}-\mu_{1}\right).
\label{eq:dMdt_H_pre}
\end{equation}
Equation~\ref{eq:dMdt_H_pre} assumes a binary gas mixture, with a light major component (atomic hydrogen) and a heavy minor component. Atmospheres are, however, composed of various species, and it is therefore necessary to define the mean molecular mass of the average effective heavy component,
\begin{equation}
    \mu_{2}=\frac{\bar{\mu}-\chi \mu_{1}}{1-\chi},
\label{eq:effective_heavy}
\end{equation}
where $\bar{\mu}$ is the local mean molecular mass. Collectively, one gets,
\begin{equation}
    \dot{M}_{1} = \frac{\chi_{\rm H}}{1{-}\chi_{\rm H}} \frac{4 \pi G M_{\rm p} \mu_{1} b_{\rm ij,H}}{k_{\rm B}T_{\rm H}}\left(\bar{\mu}-\mu_{1}\right).
\label{eq:dMdt_H}
\end{equation}
Equations~\ref{eq:dMdt_H_pre} and \ref{eq:dMdt_H} apply to planets that have achieved a diffusion-limited steady state (section~\ref{sec:steady_state}), such as Earth \citep{Joshi2019}. These equations are independent of X-ray and ultraviolet irradiation  because they describe the maximum rate at which hydrogen can be transported within the atmosphere, thus setting the limit for how much hydrogen can be lost at the top of the atmosphere. As mass loss is evaluated at the homopause, and as the homopause is neutral, the effects of photochemistry are unimportant.

\subsection{Defining the crossover mass}
\label{sec:crossover_mass}

The crossover mass is defined as the threshold mass above which particles are too heavy to be dragged along with other escaping species. Fractionation occurs when the crossover mass is low enough for a gas mixture to undergo differential separation. \citet{Hunten1987} finds the crossover mass by solving the diffusion-limited flux for $\mu_{\rm cr}$ (${=}\mu_{2}$ in Equations~\ref{eq:phi_Ht} and \ref{eq:dMdt_H_pre}),
\begin{equation}
    \mu_{\rm cr} = \mu_{1} + \frac{k_{\rm B}T \phi_{1}}{\chi b_{\rm ij}g},
\label{eq:mu_Ht}
\end{equation}
or, equivalently,
\begin{equation}
    \mu_{\rm cr} = \mu_{1} + \frac{k_{\rm B}T \dot{M}_{1}}{4 \pi \chi b_{\rm ij}GM_{\rm p}\mu_{1}}.
\label{eq:mu_Ht_2}
\end{equation}
Equations~\ref{eq:mu_Ht} and \ref{eq:mu_Ht_2} derive from Equation~\ref{eq:chapman}, which assumes that both gas species experience the same acceleration as the bulk fluid, that is, ${\rm d}v_{1}/{\rm d}t{=}{\rm d}v_{2}/{\rm d}t{=}{\rm d}v/{\rm d}t$ \citep[][p.{\,}107]{Chapman1970}. However, this assumption does not apply when evaluating fractionation because, under such circumstances, ${\rm d}v_{2}/{\rm d}t{=}0$. To address this, we use Equation~{6.62,\,9} of \citet{Chapman1970} instead, with ${\rm d}v_{2}/{\rm d}t$ in place of ${\rm d}v/{\rm d}t$:
\begin{equation}
    \rho_{2} \frac{{\rm d}v_{2}}{{\rm d}t} = -\rho_{2}g - \frac{{\rm d}P_{2}}{{\rm d}r} + \frac{\rho_{1} \rho_{2}}{\rho \tau_{1,2}}\left(v_{1}-v_{2}\right),
\label{eq:Chapman_2}
\end{equation}
where $\tau_{1,2}$ is the time between particle collisions. From Equations~{6.62,\,6} and {6.62,\,7} of \citet{Chapman1970},
\begin{equation}
    \tau_{1,2} = \frac{\rho_{1}\rho_{2}P}{P_{1}P_{2}\rho}D,
\end{equation}
which upon inserting into Equation~\ref{eq:Chapman_2} and simplifying with the ideal gas equation,
\begin{equation}
    \frac{{\rm d}v_{2}}{{\rm d}t} = -g - \frac{c_{2}^{2}}{P_{2}}\frac{{\rm d}P_{2}}{{\rm d}r} + \frac{\chi c_{2}^{2}}{D}\left(v_{1}-v_{2}\right),
\end{equation}
where $c_{2}{=}\sqrt{k_{\rm B}T/\mu_{2}}$ is the isothermal sound speed of the heavy species. Applying chain rule and dividing through by $c_{2}^{2}$,
\begin{equation}
    \frac{1}{c_{2}^{2}}\frac{{\rm d}v_{2}}{{\rm d}t} = -\frac{GM_{\rm p}\mu_{2}}{k_{\rm B}Tr^{2}} + \frac{1}{1-\chi}\frac{{\rm d}\chi}{{\rm d}r} - \frac{1}{P}\frac{{\rm d}P}{{\rm d}r} + \frac{\chi}{D}\left(v_{1}-v_{2}\right).
\label{eq:3}
\end{equation}
\citet{Hunten1973} evaluates the diffusion-limited flux at the homopause, but it is unclear if this is a suitable location for evaluating the crossover mass, which depends on local system properties. In fact, the properties of the upper atmosphere differ substantially from those in the deeper regions, and, if we are interested in knowing the compositional evolution of the bulk atmosphere, the most relevant place would be the upper edge of the atmosphere where atmospheric particles escape to space. It is thus preferable to evaluate the crossover mass at the exobase (subscript x), which is defined as the location above which gas becomes rarefied and collisions no longer dominate particle dynamics. The exobase height varies from $500{-}1000~{\rm km}$ depending on solar activity \citep{Emmert2015}; beyond the exobase, a lower crossover mass is less significant because the reduced particle density gives rise to greater statistical noise. This statistical noise invalidates the continuity assumption because individual particle motions are too significant for gas to be treated as a coherent flow \citep{Oran1998,Shematovich2015}. Evaluating Equation~\ref{eq:3} at the exobase where hydrostatic equilibrium applies \citep{Modirrousta2024} reduces the $(1/P){\rm d}P/{\rm d}r$ term to the negative inverse of the scale height of the bulk fluid, 
\begin{equation}
    \frac{1}{c_{2}^{2}}\frac{{\rm d}v_{2}}{{\rm d}t} = \frac{GM_{\rm p}\left(\bar{\mu}-\mu_{2}\right)}{k_{\rm B}T_{\rm x}R_{\rm x}^{2}} + \frac{1}{1-\chi_{\rm x}}\left(\frac{{\rm d}\chi}{{\rm d}r}\right)_{\rm x} + \frac{\chi_{\rm x}}{D_{\rm x}}\left(v_{\rm 1,x}-v_{\rm 2,x}\right).
\end{equation}
Fractionation occurs for species heavier than the crossover mass ($\mu_{2}{=}\mu_{\rm cr}$) when $v_{2}{=}{\rm d}v_{2}/{\rm d}t{=}0$ so that
\begin{equation}
    \frac{GM_{\rm p}\left(\bar{\mu}-\mu_{\rm cr}\right)}{k_{\rm B}T_{\rm x}R_{\rm x}^{2}} + \frac{1}{1-\chi_{\rm x}}\left(\frac{{\rm d}\chi}{{\rm d}r}\right)_{\rm x} + \frac{\chi_{\rm x} v_{\rm 1,x}}{D_{\rm x}} = 0.
\end{equation}
Multiplying the numerator and denominator of the rightmost term by $n$ yields,
\begin{equation}
    \frac{GM_{\rm p}\left(\bar{\mu}-\mu_{\rm cr}\right)}{k_{\rm B}T_{\rm x}R_{\rm x}^{2}} + \frac{1}{1-\chi_{\rm x}}\left(\frac{{\rm d}\chi}{{\rm d}r}\right)_{\rm x} + \frac{\phi_{\rm x}}{b_{\rm ij,x}} = 0,
\end{equation}
which can also be expressed in terms of the mass flow rate,
\begin{equation}
    \frac{GM_{\rm p}\left(\bar{\mu}-\mu_{\rm cr}\right)}{k_{\rm B}T_{\rm x}R_{\rm x}^{2}} + \frac{1}{1-\chi_{\rm x}}\left(\frac{{\rm d}\chi}{{\rm d}r}\right)_{\rm x} + \frac{\dot{M}_{1}}{4 \pi R_{\rm x}^{2} \mu_{1} b_{\rm ij,x}} = 0.
\end{equation}
The compositional term is small and can be discarded \citep{Zahnle1986}. Solving for $\mu_{\rm cr}$,
\begin{equation}
    \mu_{\rm cr} = \bar{\mu} + \frac{k_{\rm B}T_{\rm x} \dot{M}_{1} }{4 \pi b_{\rm ij,x} GM_{\rm p} \mu_{1}},
\label{eq:mu_dM/dt}
\end{equation}
where we see that Equations~\ref{eq:mu_Ht_2} and \ref{eq:mu_dM/dt} share a similar functional form but with different parameter values, and $\chi$ not being in the denominator. When a diffusion-limited steady state applies (section~\ref{sec:steady_state}), the hydrogen mass flow rate, $\dot{M}_{1}$, is set by Equation~\ref{eq:dMdt_H} and cannot be replaced by alternative models such as the energy-limited \citep{Watson1981} or hydro-based \citep{Kubyshkina2018(1)} approximations (discussed in section~\ref{sec:discussion}). As mentioned previously, diffusion-limited mass loss combines advection and momentum diffusion. This interaction creates a negative feedback mechanism in multicomponent gas mixtures where extreme mass loss becomes unsustainable because heavy species slow down escaping hydrogen, reducing the crossover mass, and leading to increased fractionation and further decreased hydrogen outflow. It would therefore be misleading to use an extreme mass loss model that assumes no fractionation because this is the very reason why they exhibit extreme mass loss in the first place. Such models inherently preclude fractionation because they do not incorporate the drag effect imposed by heavy species on escaping hydrogen. To self-consistently quantify fractionation in a multicomponent system, we therefore use Equation~\ref{eq:dMdt_H} and combine it with $b_{\rm ij}{\approxprop}T^{3/4}$ \citep{Mason1970,Marrero1972},
\begin{equation}
    \mu_{\rm cr} = \bar{\mu}_{\rm x} + \frac{\chi_{\rm H}}{1{-}\chi_{\rm H}}\left(\frac{T_{\rm x}}{T_{\rm H}}\right)^{\frac{1}{4}}\left(\bar{\mu}_{\rm H}-\mu_{1}\right).
\label{eq:mu_MGK}
\end{equation}
For well-mixed gas, such as exoplanets with primordial atmospheres \citep[i.e., $K_{\rm zz}{\gg}D$;][]{Modirrousta2023}, $\bar{\mu}_{\rm x}{\approx}\bar{\mu}_{\rm H}{\approx}\bar{\mu}$ and,
\begin{equation}
    \mu_{\rm cr} = \bar{\mu}\left[1 + \frac{\chi_{\rm H}}{1{-}\chi_{\rm H}}\left(\frac{T_{\rm x}}{T_{\rm H}}\right)^{\frac{1}{4}}\left(1-\frac{\mu_{1}}{\bar{\mu}}\right)\right].
\label{eq:mu_MGK2}
\end{equation}
Equations~\ref{eq:mu_MGK} and \ref{eq:mu_MGK2} apply to all planets with compositionally stratified and well-mixed ideal gas atmospheres, respectively, that have achieved a diffusion-limited steady state (section~\ref{sec:steady_state}). They suggest that in the diffusion-limited mass loss framework, the crossover mass is independent of the hydrogen outflow rate but depends on the hydrogen mole fraction, the temperature ratio to the one-fourth power, and the atmospheric composition. The photosphere (where optical depth $\tau{=}2/3$) is a suitable alternative to the homopause for exoplanets with hydrogen-rich atmospheres because it is also a temperature minimum \citep[i.e., ${\rm d}T/{\rm d}r{=}0$;][]{Modirrousta2023}. In the following section, we compare our approach to that of \citet{Hunten1987}.

\subsection{Model comparison}
\label{sec:model_comparison}

\begin{figure}[htbp]
    \centering
    \includegraphics[width=0.6\linewidth]{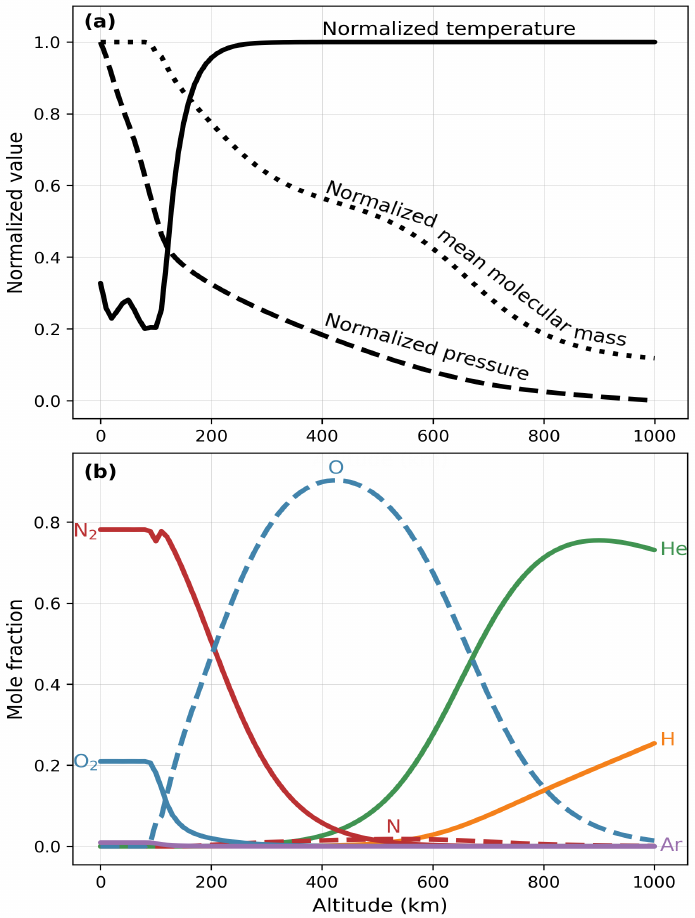}
    \caption{(a) Normalized temperature (solid line), mean molecular mass (dotted line), and pressure (dashed line). The temperature, mean molecular mass, and pressure are normalized with $T/\left(923.5~{\rm K}\right)$, $\bar{\mu}/28.96$, and $\log_{10}\left(P/P_{\rm min}\right)/\log_{10}\left(P_{\rm max}/P_{\rm min}\right)$, respectively, where $P_{\rm max}{=}1.002{\times}10^{5}~{\rm Pa}$ and $P_{\rm min}{=}2.969{\times}10^{-9}~{\rm Pa}$ are the maximum and minimum pressures in our atmospheric grid. (b)~The terrestrial compositional profile for hydrogen (H, solid orange line), helium (He, solid green line), oxygen (blue dashed line for O and blue solid line for O$_{2}$), nitrogen (red dashed line for N and red solid line for N$_{2}$), and argon (Ar, solid purple line). Data from the NRLMSIS 2.0 code \citep{Emmert2021}.}
    \label{fig:profiles}
\end{figure}

Our model for the crossover mass (Equations~\ref{eq:mu_MGK} and \ref{eq:mu_MGK2}) and that of \citet{Hunten1987} (Equations~\ref{eq:mu_Ht} and \ref{eq:mu_Ht_2}) are based on different assumptions. We compare their predictions with the observations of Earth's atmosphere, using the NRLMSIS~2.0 code \citep{Emmert2021}, which provides the average observed behavior of an atmospheric column at a given latitude, longitude, and elevation. Athens, Greece (37.9838$^{\circ}${\,}N, 23.7275$^{\circ}${\,}E) is chosen as the reference location because of its proximity to multiple continents and its mean annual temperature being similar to Earth's average temperature (${\sim}288~{\rm K}$). From the output data, we extract the mole fraction of hydrogen (H), helium (He), oxygen (O and O$_{2}$), nitrogen (N and N$_{2}$), and argon (Ar), along with the mean molecular mass, pressure, and temperature for elevations of $0{-}1000~{\rm km}$. Figure~\ref{fig:profiles} shows the parameter and compositional profiles adopted in this study.

\begin{figure}[htbp]
    \centering
    \includegraphics[width=0.7\linewidth]{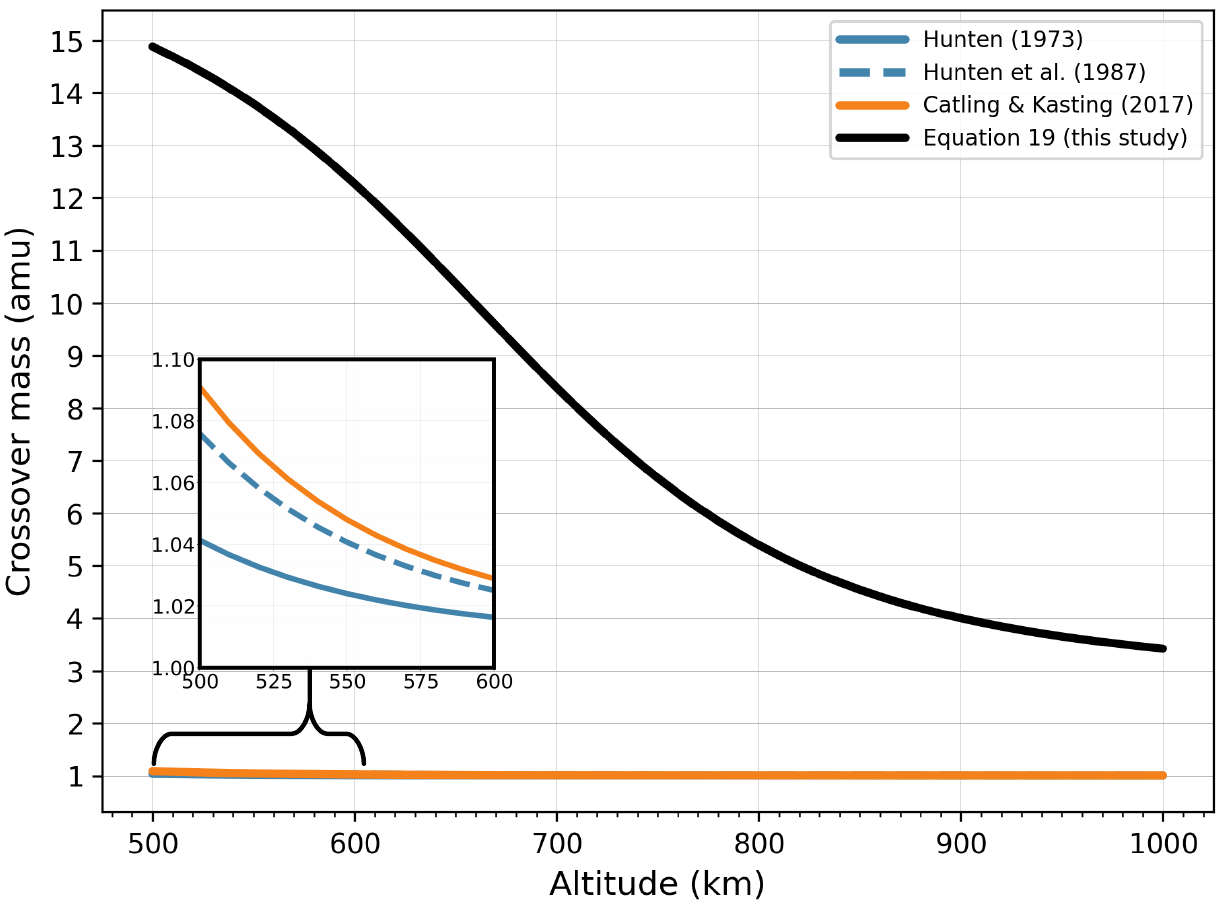}
    \caption{The crossover mass at various altitudes at the exobase. The black line is our suggested crossover mass function (Equation~\ref{eq:mu_MGK}) while the colored lines correspond to Equation~\ref{eq:mu_Ht} with the following binary diffusion coefficients: $b_{\rm ij}{=}2.67{\times}10^{19}T^{0.75}~{\rm m^{-1}~s^{-1}}$ \citep[blue solid line;][]{Hunten1973}, $b_{\rm ij}{=}2.2{\times}10^{21}~{\rm m^{-1}~s^{-1}}$ \citep[blue dashed line;][]{Hunten1987}, and $b_{\rm ij}{=}1.8{\times}10^{21}~{\rm m^{-1}~s^{-1}}$ \citep[orange line;][p.{\,}146]{Catling2017}.}
    \label{fig:crossover_mass}
\end{figure}

Equations~\ref{eq:mu_Ht} and \ref{eq:mu_MGK} are applied to the exobase ($500{-}1000~{\rm km}$), with the former using the binary diffusion coefficients of \citet{Hunten1973}, \citet{Hunten1987}, and \citet{Catling2017}. We obtain the required particle flux, $\phi_{1}$, using the parameterization $\phi_{1} {=} 1.2 {\times} 10^{12}\left(r/6837~{\rm km}\right)^{-2}~{\rm m^{-2}~s^{-1}}$, which derives from observations of Earth's thermosphere \citep{Joshi2019}. Figure~\ref{fig:crossover_mass} shows that the crossover mass of Equation~\ref{eq:mu_Ht} remains relatively constant at around $1~{\rm amu}$, with little variation across different binary diffusion coefficient values. In contrast, our crossover mass formulation (Equation~\ref{eq:mu_MGK}) yields a value of $9{\pm}6~{\rm amu}$, and varies with altitude because of the changing mean molecular mass across the heterosphere, suggesting that fractionation occurs primarily at higher altitudes.

The crossover mass value we obtain when using the approach of \citet{Hunten1987}, $\mu_{\rm cr}{\approx}1~{\rm amu}$, differs from the value given in their study, $\mu_{\rm cr}{=}2.5~{\rm amu}$, and in other iterations of their work \citep[e.g., $\mu_{\rm cr}{=}2.25~{\rm amu}$;][Table~5.3]{Catling2017}. This discrepancy can be understood through differences in the assumptions and parameter values adopted when evaluating the crossover mass. First, these studies derive the crossover mass from Equation~\ref{eq:chapman}, which assumes that all gas species have equal acceleration \citep[][p.{\,}107]{Chapman1970}. This is physically inconsistent because fractionation can occur only when different species accelerate at different rates and, thus, Equation~\ref{eq:dMdt_H} should be used instead. Second, the energy-limited approximation \citep{Watson1981} is used to estimate hydrogen flux whereas we use present-day observations of leaking hydrogen \citep{Joshi2019}. The use of the energy-limited approximation is inadequate because it assumes free advection, making it inherently incapable of evaluating fractionation. Fractionation involves differential separation resulting from momentum diffusion exchange between different species, which is not incorporated in the energy-limited approximation. Third, they assume a hydrogen mole fraction close to one ($\chi{\approx}1$), which is valid only for hydrogen-rich atmospheres and not for the present-day Earth (Figure~\ref{fig:profiles}).

\begin{figure}[htbp]
    \centering
    \includegraphics[width=0.7\linewidth]{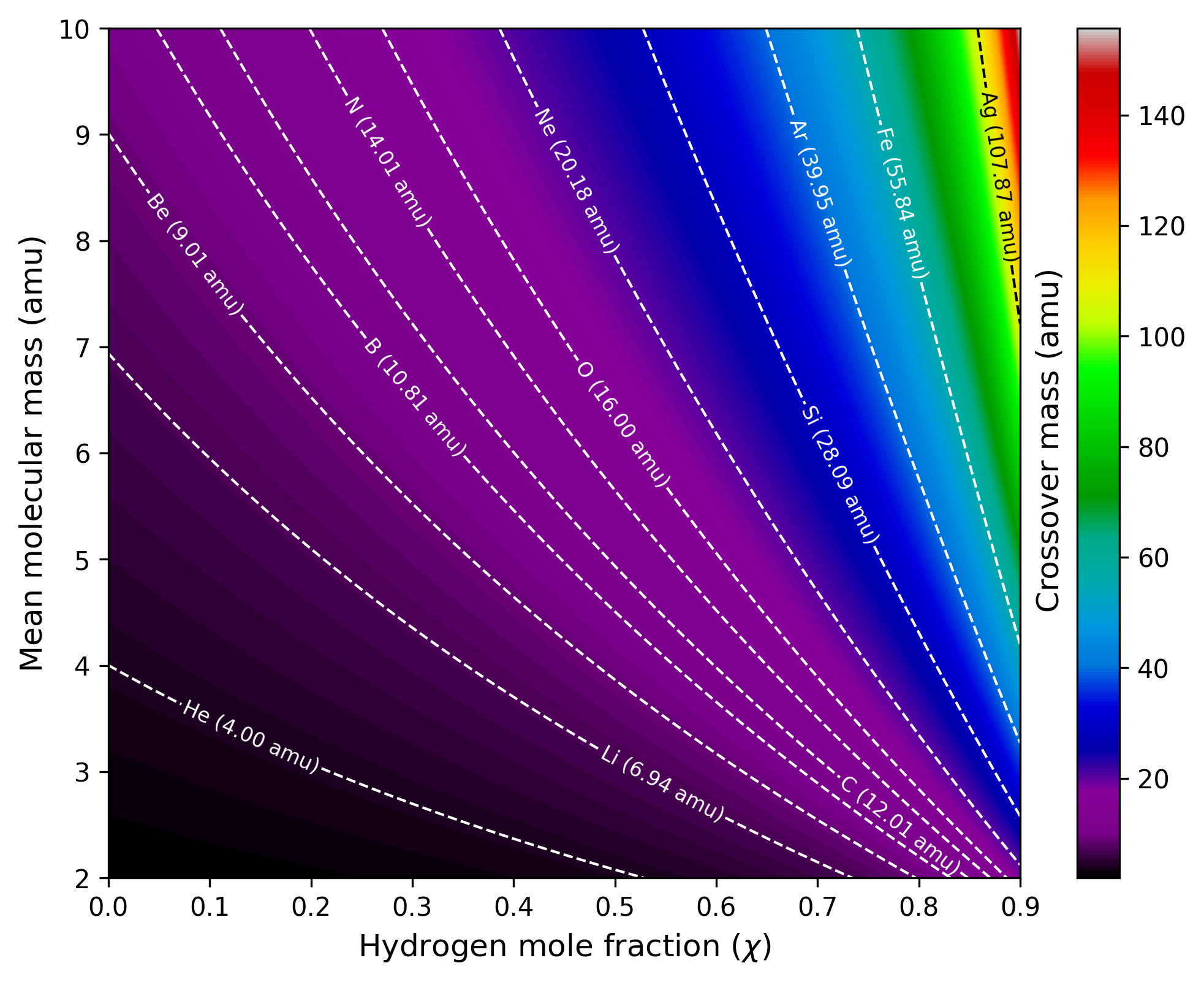}
    \caption{The crossover mass (Equation~\ref{eq:mu_MGK2}) as a function of the mean molecular mass and the hydrogen mole fraction for $T_{\rm x}/T_{\rm H}{=}10$. The contours show the regions at which various elements are fractionated.}
    \label{fig:crossover_mass_2}
\end{figure}

In the context of exoplanets with primordial atmospheres, Equation~\ref{eq:mu_MGK2} can be used because the hydrogen mole fraction is high across the entire atmosphere ($\chi{\approx}0.9$) and mixing from eddy diffusion is efficient throughout \citep{Parmentier2013,Charnay2015}. If compositional stratification occurs, Equation~\ref{eq:mu_MGK} should be employed at the top of the atmosphere (i.e., the exobase) instead. Equation~\ref{eq:mu_MGK} is independent of the chemical gradient in the deeper sections of the atmosphere because the crossover mass at the exobase sets the fractionation bottleneck for the entire atmosphere. Figure~\ref{fig:crossover_mass_2} shows that the crossover mass increases with $\chi$, suggesting that the preferential loss of hydrogen leads to a progressively more stable (i.e., more fractionating) and hydrostatic atmosphere composed of heavier species.

\section{Application to exoplanets}
\subsection{Rapid onset of steady state}
\label{sec:steady_state}

In the previous section we discussed how planets with hydrogen-rich upper-atmospheres are prone to compositional fractionation because the low crossover mass prevents heavy species from escaping, leading to local hydrogen depletion. However, information about this depletion is not felt instantaneously across the atmosphere because only the region immediately below this depleted layer can feel the effect of the compositional gradient. It is for this reason that the diffusion limit applies only when hydrogen flow is uniform across all atmospheric spherical shells, and only through achieving such a steady state does mass flow become limited by the maximum rate at which it can be transported within the atmosphere.

\begin{figure}[htbp]
    \centering
    \includegraphics[width=1\linewidth]{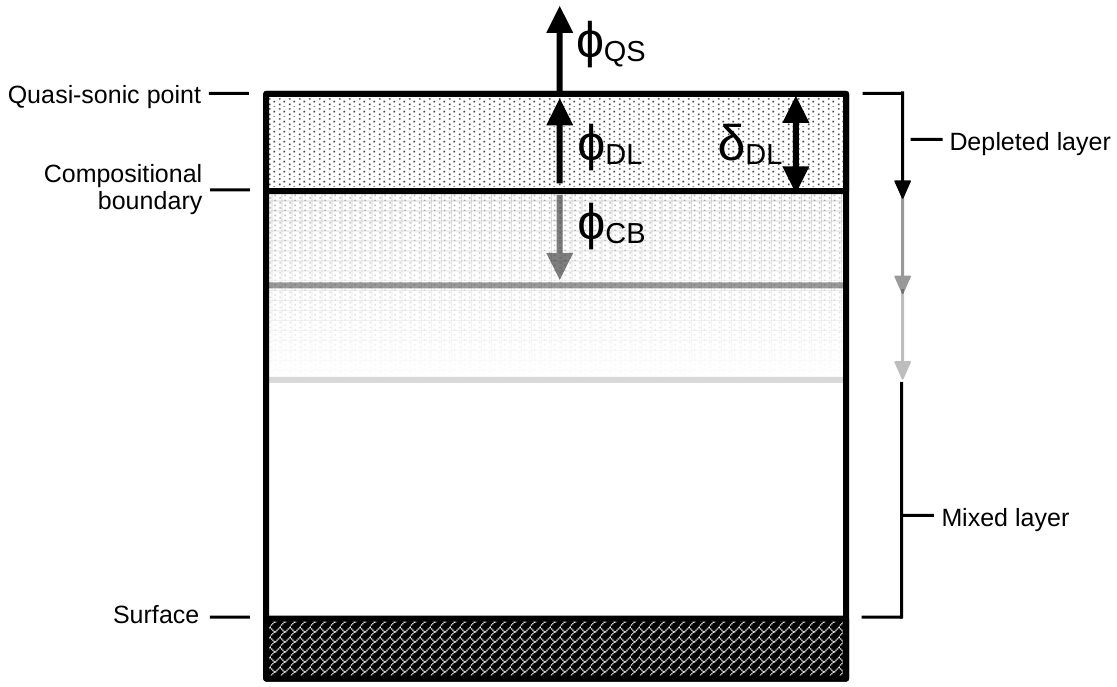}
    \caption{Cartoon showing the assumed structure of the upper atmosphere after developing a depleted layer. Continuity requires the hydrogen flux at the quasi-sonic point, $\phi_{\rm QS}$, to equal that through the depleted layer, $\phi_{\rm DL}$, and the growing compositional boundary $\phi_{\rm CB}$. Diagram not to scale.}
    \label{fig:cartoon}
\end{figure}

In this section, we explore and apply this concept to exoplanets with primordial envelopes experiencing photoevaporation. For mass loss to continue, there must be a preferential transport of hydrogen from deeper layers; otherwise the mole fraction of hydrogen, $\chi$, approaches zero, and mass loss ceases. This transport can occur either through advection or diffusion. If the mechanism is advection, the upper atmosphere will become exceedingly more enriched as more hydrogen is lost and more heavier species are left behind. In other words, the combination of preferential hydrogen removal at the top of the atmosphere and advection with the average composition would lead to further enrichment, preventing a steady state. Therefore, if advection were the restocking mechanism, the concentration gradient would keep growing until diffusive transport restores equilibrium. Consequently, a steady state can be reached only by balancing the preferential loss of hydrogen with the preferential restocking of hydrogen, which can occur only through diffusion. The balance between diffusion and advection determines whether chemical inhomogeneities form. If diffusion is efficient, these inhomogeneities will disperse rapidly. To evaluate whether this scenario is possible, we consider a hydrogen-rich hydrodynamic atmosphere that has just experienced preferential hydrogen loss. We follow the framework of \citet{Modirrousta2024}, where it was shown that the exobase and quasi-sonic point coincide for hydrodynamic atmospheres, and we therefore refer to the highest point in the atmosphere as the quasi-sonic point (subscript QS) because we assume the atmosphere is initially hydrodynamic. This point represents the velocity maximum in the atmosphere, which, in the limit of $v{=}c_{\rm s}$, corresponds to the sonic point. Figure~\ref{fig:cartoon} is a schematic diagram of the configuration being examined.

After experiencing mass loss, diffusion will act on the locally formed compositional gradient within a timescale of,
\begin{equation}
    t_{\rm df} = \frac{\delta_{\rm DL}^{2}}{2\left(K_{\rm zz}{+}D\right)},
\end{equation}
where $\delta_{\rm DL}$ is the depleted layer depth. Assuming the atmosphere begins in a highly hydrodynamic state, eddy diffusion greatly exceeds molecular diffusion (i.e., $K_{\rm zz}{\gg}D$). The eddy diffusion coefficient is defined as $K_{\rm zz}{=}\langle v L_{\rm m} \rangle$ where $v{=}v_{\rm QS}$ is the bulk vertical wind velocity at the quasi-sonic point and $L_{\rm m}$ is the mixing length. The mixing length is not well understood because it is subject to highly nonlinear dynamical processes, such as gravitational wave breaking \citep[e.g.,][]{Lindzen1971,Lindzen1981}. Experiments suggest that $L_{\rm m}{\sim}C_{\rm vK}L_{\rm c}$, where $C_{\rm vK}{\approx}0.4$ is the von Kármán constant \citep{Pope2000} and $L_{\rm c}$ is the characteristic length, which is sensitive to local system properties. In this context, the characteristic length is the depleted layer size, obtaining $K_{\rm zz}{\sim}\langle C_{\rm vK}v_{\rm QS}\delta_{\rm DL}\rangle$. Evaluating the above
\begin{equation}
    t_{\rm df} > \frac{5\delta_{\rm DL}}{4v_{\rm QS}} = \frac{5}{4}t_{\rm a},
\end{equation}
where $t_{\rm a}$ is the advection timescale, demonstrating that diffusion is less efficient than advection. Under such circumstances, a locally depleted layer must form because advection removes hydrogen faster than it is restocked. This will cause the hydrogen mole fraction at the quasi-sonic point to decrease, creating a depleted layer through which hydrogen diffuses. As mass loss continues, the hydrogen content in the region immediately below the depleted layer diffuses through it and is subsequently lost at the quasi-sonic point. The size of the depleted layer will therefore increase with further mass loss (Figure~\ref{fig:cartoon}). 
This gives rise to three equations for mass loss: one for advection at the quasi-sonic point,
\begin{equation}
    \phi_{\rm QS} = \chi_{\rm QS} n_{\rm QS} v_{\rm QS},
\label{eq:phi_B}
\end{equation}
one for hydrogen diffusion through the depleted layer,
\begin{equation}
    \phi_{\rm DL} = -\left(\bar{K}_{\rm zz}+\bar{D}\right)\frac{\chi_{\rm QS} n_{\rm QS}-\chi_{\rm CB}n_{\rm CB}}{\delta_{\rm DL}},
\label{eq:phi_DL}
\end{equation}
and the other for the growth rate of the depleted layer,
\begin{equation}
    \phi_{\rm CB} = \chi_{\rm CB} n_{\rm CB} \frac{{\rm d}\delta_{\rm DL}}{{\rm d}t}.
\label{eq:phi_CB}
\end{equation}
The parameters $\chi_{\rm CB}$ and $n_{\rm CB}$ are the mole fraction of hydrogen and the volumetric total particle number density at the bottom of the depleted layer (labeled the compositional boundary), and $\bar{K}_{\rm zz}$ and $\bar{D}$ are the average eddy and molecular diffusion coefficients,
\begin{equation}
    \bar{K}_{\rm zz}{+}\bar{D} = \frac{1}{R_{\rm QS}-R_{\rm CB}\left(t\right)}\int^{R_{\rm QS}}_{R_{\rm CB}\left(t\right)} K_{\rm zz}\left(r\right) {+} D\left(r\right){\rm d}r.
\end{equation}
The above equations can be solved analytically if a plane-parallel atmosphere and constant average diffusion coefficients are assumed,
\begin{equation}
    \bar{K}_{\rm zz}{+}\bar{D} \approx \frac{1}{R_{\rm QS}-R_{0}}\int^{R_{\rm QS}}_{R_{0}} K_{\rm zz}\left(r\right) {+} D\left(r\right){\rm d}r,
\end{equation}
with the lower limit set to the planetary surface ($R_{0}$). Balancing Equations~\ref{eq:phi_DL} and \ref{eq:phi_CB},
\begin{equation}
    \delta_{\rm DL} \frac{{\rm d}\delta_{\rm DL}}{{\rm d}t} = \left(\bar{K}_{\rm zz}+\bar{D}\right)\left(1-\frac{\chi_{\rm QS} n_{\rm QS}}{\chi_{\rm CB}n_{\rm CB}} \right).
\label{eq:delta_ddelta_dt}
\end{equation}
The mole fraction and number density ratio is found by balancing Equations~\ref{eq:phi_B} and \ref{eq:phi_DL},
\begin{equation}
    \frac{\chi_{\rm QS} n_{\rm QS}}{\chi_{\rm CB}n_{\rm CB}}= \frac{\bar{K}_{\rm zz}+\bar{D}}{v_{\rm QS}\delta_{\rm DL}+\bar{K}_{\rm zz}+\bar{D}},
\end{equation}
which can be inserted into Equation~\ref{eq:delta_ddelta_dt} and integrated,
\begin{equation}
    \int^{R_{\rm QS}{-}R_{0}}_{0} \left(\frac{\delta_{\rm DL}}{\bar{K}_{\rm zz}+\bar{D}} + \frac{1}{v_{\rm QS}}\right){\rm d}\delta_{\rm DL} = \int^{t}_{0} {\rm d}t,
\label{eq:integral}
\end{equation}
yielding,
\begin{equation}
\begin{split}
    t &= \frac{\left(R_{\rm QS}-R_{0}\right)^{2}}{2\left(\bar{K}_{\rm zz}+\bar{D}\right)}+\frac{R_{\rm QS}-R_{0}}{v_{\rm QS}} \\
    &\approx \frac{\left(R_{\rm QS}-R_{0}\right)^{2}}{2{K}_{\rm zz}}
\end{split}
\label{eq:time}
\end{equation}
Evaluating Equation~\ref{eq:time} for typical values relevant to Earth's atmosphere, with $\bar{K}_{\rm zz}{\sim}10^{2}~{\rm m^{2}~s^{-1}}$ \citep{Liu2021}, suggests that the time required for steady state to apply is geologically negligible (${\lesssim}40~{\rm years}$). In fact, general circulation atmospheric models suggest that exoplanets may possess significantly higher eddy diffusion coefficients because they reside in more extreme environments \citep[e.g.,][]{Parmentier2013,Charnay2015}, indicating that steady state may be established in even shorter timescales. The amount of atmospheric depletion that occurs during the onset of steady state depends on the hydrogen outflow rate at the quasi-sonic point. Even with inviscid XUV-induced photoevaporation, the overall impact is likely negligible because total mass loss requires $10^{3}{-}10^{5}$ years (Figure~\ref{fig:mass_loss_evolution}). Higher mass loss rates, however, may result in complete atmospheric loss during the steady-state onset, and we discuss this possibility in section~\ref{sec:self_consistent}.

\subsection{Representative simulations}
\label{sec:simulations}

Having assessed the rapid onset of steady state in planets with hydrogen-rich atmospheres, we now present several representative simulations for exoplanets experiencing mass loss. Because hydrogen loss at the quasi-sonic point has to be supplied from below, diffusion-limited mass loss at any point is constrained by the region just below it. The global diffusion-limited mass loss rate is therefore equal at all points in the atmosphere. Unlike \citet{Hunten1973} and \citet{Hunten1987}, we evaluate mass loss at the photosphere (subscript p; where optical depth $\tau{=}2/3$) because it is the temperature minimum of the atmosphere and because the compositional gradient is small \citep[$K_{\rm zz}{\gg}D$;][]{Modirrousta2023}, allowing us to use Equation~\ref{eq:dMdt_H}. We replace the experimentally-derived binary diffusion coefficient, $b_{\rm ij}$, with the more general molecular diffusion coefficient and photospheric number density, $n_{\rm p}D$. To estimate the diffusion coefficient we apply the Chapman-Enskog hard sphere approximation \citep{Chapman1970} with the average particle kinetic radius taken as that of an atomic hydrogen-helium mixture ($\sigma_{\rm 12}{=}2{\times}10^{-10}~{\rm m}$),
\begin{equation}
\begin{split}
    \dot{M}_{1} &= 1.1{\times}10^{5}\frac{\chi_{\rm p}}{1{-}\chi_{\rm p}}\left(\frac{\bar{\mu}_{\rm p}}{\mu_{1}}{-}1 \right)\left(\frac{\mu_{1}}{\mu_{2}}+1\right)^{\frac{1}{2}}\left(\frac{M_{\rm p}}{M_{\oplus}}\right)\left(\frac{T_{\rm p}}{1000~{\rm K}}\right)^{-\frac{1}{2}}~{\rm kg~s^{-1}} \\
    & \approx 1.1{\times}10^{5}\frac{\chi_{\rm p}}{1{-}\chi_{\rm p}}\left(\frac{\bar{\mu}_{\rm p}}{\mu_{1}}{-}1 \right)\left(\frac{M_{\rm p}}{M_{\oplus}}\right)\left(\frac{T_{\rm p}}{1000~{\rm K}}\right)^{-\frac{1}{2}}~{\rm kg~s^{-1}},
\end{split}
\label{eq:our_model_2}
\end{equation}
where $\left(\mu_{1}/\mu_{2}{+}1\right)^{1/2}{\approx}1$. Simulations suggest that eddy diffusion is consistently greater than molecular diffusion in hydrogen-rich atmospheres \citep{Modirrousta2023}, so $\chi_{\rm p}$ can be approximated with $M_{\rm H}/\left(M_{\rm H}{+}M_{\rm z}\right)$, where $M_{\rm H}$ and $M_{\rm z}$ are the total atmospheric hydrogen and heavy components. For non-gas giant planets, the heavy component originates mainly from outgassed carbon dioxide during magma ocean solidification \citep{Elkins2008(1),Lebrun2013,Salvador2017,Bower2019}, with water being released more gradually because of its high solubility in magma \citep{Lichtenberg2021,Miyazaki2022}. The initial total non-hydrous volatile budget of silicate planets is thought to be of the order ${\sim}0.01\%$ by weight \citep{Hirschmann2009,Marty2012} so that $M_{\rm z}{\sim}10^{-4}M_{\rm p}$. The initial accreted primordial atmosphere is less well known, with simulations providing a wide range of estimates \citep{Lee2015,Ginzburg2016,Mordasini2020}, so we set it as a free parameter. We therefore approximate the mean molecular mass as,
\begin{equation}
    \bar{\mu}=\max{\left[2.2~{\rm amu}, \chi_{\rm p}\mu_{\rm H}{+}(1{-}\chi_{\rm p})\mu_{\rm CO_{2}}\right]},
\end{equation}
with $\mu_{\rm H}$ and $\mu_{\rm CO_{2}}$ being the molecular masses of atomic hydrogen and carbon dioxide. Our test planet is GJ~357~b, which is a super-Earth with a mass, radius, and equilibrium temperature of $1.84{\pm}0.31~M_{\oplus}$ \citep{Luque2019}, $1.20{\pm}0.06~R_{\oplus}$ \citep{Oddo2023}, and $525~{\rm K}$ \citep{Luque2019} respectively. The X-ray luminosity and age of the host star were determined through XMM-Newton observations to be $L_{\rm x}{=}10^{18.73}~{\rm W}$ and $5~{\rm Gyr}$ \citep{Modirrousta2020c}. The X-ray and ultraviolet luminosity evolution of the star are estimated through empirical relations \citep[][see also \citealt{Chadney2015,Johnstone2021}]{Penz2008(2),Sanz-Forcada2011}, which we extend back to right after the T Tauri stage (age${\sim}10~{\rm Myr}$). For the atmospheric size we employ the model of \citet{Lopez2014}, which provides the atmospheric radius as a function of the planetary mass, the atmospheric mass fraction, the equilibrium temperature, and the planetary age. To provide a basis for comparison with the diffusion-limited mass loss model, we also use the energy-limited approximation of \citet{Watson1981} and the hydro-based model of \citet{Kubyshkina2018(1)}. Both models assume inviscid free advection, but the latter considers thermal heating and chemistry, usually resulting in higher mass loss rates.
\begin{figure}[htbp]
    \centering
    \includegraphics[width=0.8\linewidth]{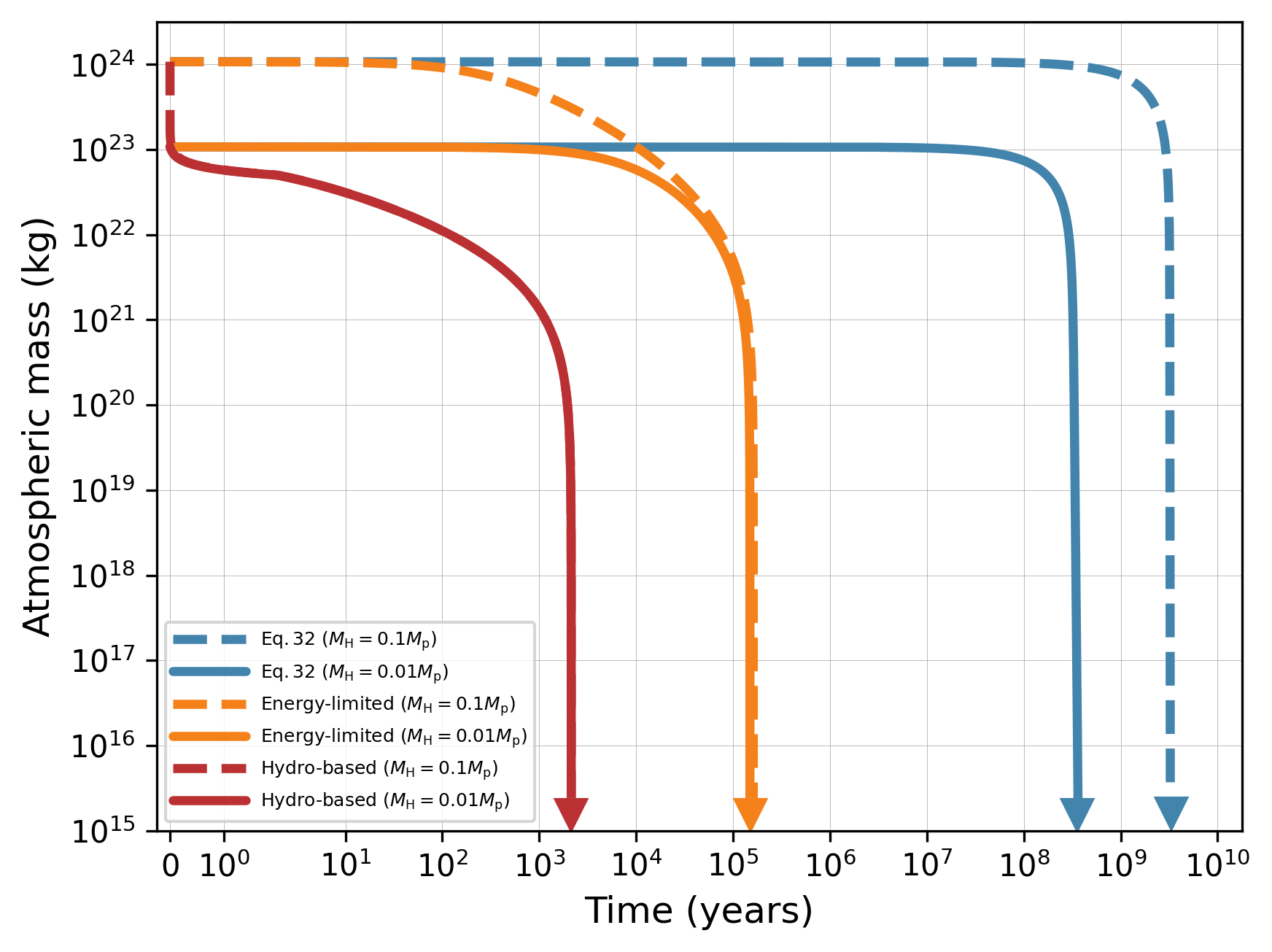}
    \caption{Atmospheric mass as a function of time for super-Earth GJ~357~b ($M_{\rm p}{=}1.84{\pm}0.31~M_{\oplus}$, $R_{\rm p}{=}1.20{\pm}0.06~R_{\oplus}$, $T_{\rm eq}{=}525~{\rm K}$, $L_{\rm x}{=}10^{18.73}~{\rm W}$) modeled using the hydro-based (red line), energy-limited (orange line), and diffusion-limited (blue line) mass loss models. The solid and dashed lines are for initial atmospheric hydrogen budgets of $M_{\rm H}{=}0.01M_{\rm p}$ and $0.1M_{\rm p}$ respectively.}
    \label{fig:mass_loss_evolution}
\end{figure}

Figure~\ref{fig:mass_loss_evolution} shows the mass loss evolution of GJ~357~b for initial atmospheric hydrogen budgets of $M_{\rm H}{=}0.1M_{\rm p}$ and $0.01M_{\rm p}$. The diffusion-limited model predicts lower mass loss rates than the hydro-based and energy-limited models, enabling the retention of atmospheric hydrogen over significantly longer timescales. Assuming the stellar age of $5~{\rm Gyr}$ is coeval with that of the planet \citep{Modirrousta2020c}, GJ~357~b probably accreted an atmosphere less than ${\sim}0.16~M_{\rm p}$; otherwise, it would have survived photoevaporation and still existed. This model, therefore, provides tighter constraints on the mass loss evolution of GJ~357~b than the hydro-based model, which predicts an initial hydrogen reservoir less than ${\sim}21~{\rm M_{p}}$ \citep{Modirrousta2020c}. 

\section{Discussion}
\label{sec:discussion}

\subsection{On the self-consistency of mass loss and fractionation}
\label{sec:self_consistent}

Having evaluated diffusion-limited mass loss (section~\ref{sec:diffusion_limited}), the crossover mass (section~\ref{sec:crossover_mass}), and the transition to the diffusion-limited regime (section~\ref{sec:steady_state}), we now discuss the self-consistency of mass loss and fractionation. 

In \citet{Modirrousta2023}, atmospheric evaporation is categorized into three regimes. In the first regime, mass loss arises from internal energy generated during accretion; in the second regime, mass loss occurs through a combination of internal energy and incoming thermal radiation; and in the third regime, mass loss is driven by X-ray and ultraviolet irradiation. Selecting the appropriate mass loss regime depends on the thermodynamic properties of the system, which we explored in our previous paper. To keep our explanation concise and avoid reiterating what has already been discussed in \citet{Modirrousta2023}, we focus only on regimes one and three, which represent endmember cases when mass loss is driven by either internal energy or incoming X-ray and ultraviolet irradiation.

Regime one (also known as core-powered mass loss) is very efficient and can remove a primordial atmosphere in a few days \citep{Modirrousta2023}, which is too fast for steady state to activate (${\lesssim} 40~{\rm years}$) and diffusion-limited mass loss to apply. This suggests that fractionation is unlikely to occur as a result of geological events that bring a planet into regime one, such as giant impacts \citep[e.g.,][]{Canup2008,Lock2018} or late accretion \citep[e.g.,][]{Marchi2018,Marchi2023}. In contrast, regime three (also known as XUV photoevaporation) proceeds more slowly, requiring $10^{3}{-}10^{5}$ years to remove an atmosphere from super-Earth and sub-Neptune exoplanets (Figure~\ref{fig:mass_loss_evolution}). These longer timescales allow for steady state to activate and diffusion-limited mass loss to apply. Consequently, whereas a planet in regime three may initially experience mass loss according to traditional inviscid free advection models (e.g., the energy-limited and the hydro-based models), it will rapidly shift to diffusion-limited mass loss after the onset of steady state.

\subsection{On modeling photoevaporation accurately}

Of the three regimes of atmospheric evaporation, the third is the most widely discussed because of its prolonged duration and observability through Lyman~$\alpha$ spectroscopy \citep{Linsky2010,Lecavelier2012,Vidal2013,Rockcliffe2023}. The prevailing model employed for assessing mass loss in the third regime is the energy-limited model \citep{Watson1981}. Despite its apparent simplicity and intuitive nature, this model is beset by several well-documented limitations \citep[e.g.,][]{Kubyshkina2018(1),Krenn2021,Modirrousta2023}. Most notably, it neglects the influence of incoming thermal radiation and erroneously assumes that incoming XUV energy is deposited at the XUV-photosphere ($\tau_{\rm XUV}{=}2/3$). These assumptions are hard to justify because exoplanets are often highly irradiated, and in the case of primordial atmospheres, the XUV photosphere is usually above the sonic point \citep{Sekiya1980,Sekiya1981}. Therefore, the sonic point \citep[or more correctly, the quasi-sonic point;][]{Modirrousta2024} is a better location for assessing XUV energy deposition.

To overcome these limitations, \citet{Kubyshkina2018(1)} proposed the Hydro-based model, which incorporates a more realistic chemical analysis and addresses the limitations of the energy-limited approximation. However, this model also makes several assumptions that are likely to be invalid. First, they assume that gas accelerates indefinitely, which is not always accurate \citep{Modirrousta2024}. Second, it assumes that mass loss occurs through inviscid free advection for prolonged durations, and it is therefore efficient in stripping hydrogen from planetary atmospheres. Consequently, the question still remains as to why some exoplanets maintain their hydrogen-rich atmospheres while others do not. If, indeed, fractionation is significant, this could provide a plausible explanation for the observed diversity in exoplanet atmospheres because extreme hydrogen loss is unsustainable if fractionation takes place and mass loss becomes diffusion-limited. In other words, the energy-limited and hydro-based models are both limited in their ability to describe mass loss, and they should be applied judiciously. In contrast, the diffusion-limited mass loss model incorporates the effects of chemical fractionation, and it therefore provides a more realistic description of mass loss. 

\subsection{On the observability of chemical fractionation in exoplanetary atmospheres}

Our model suggests that planetary atmospheres become chemically fractionated because of XUV-induced photoevaporation, transitioning from a primordial hydrogen-rich atmosphere to a secondary heavier one. The specific details of this evolution depend on the planet's bulk properties and its environment, requiring a case-by-case evaluation. This evolution may be detectable using atmospheric spectroscopy \citep{Burrows2014,Madhusudhan2019}. However, observations are limited to cloudless planets because clouds obstruct visibility into deeper atmospheric layers \citep{Betremieux2017,Betremieux2018} and create pronounced chemical stratification, likely masking the effects of diffusion-limited chemical fractionation. The recent launch of the James Webb Space Telescope and upcoming missions like Ariel and Twinkle hold promise for advancing our understanding of the link between chemical fractionation and planetary evolution.

\section{Concluding Remarks}

In this study, we suggest that X-ray and ultraviolet-induced photoevaporation in primordial atmospheres results in compositional fractionation. When studying Earth, unlike earlier research, we assess fractionation at the exobase. We find that the crossover mass ranges from $4{-}15~{\rm amu}$ for Earth, with it depending primarily on the altitude at which it is evaluated, the hydrogen mole fraction, the temperature ratio to the one-fourth power, and the local mean molecular mass. Building on this result, we demonstrate that mass fractionation results in a transient disequilibrium, forming a hydrogen-depleted layer that grows from the top of the atmosphere downward. After reaching steady state, hydrodynamic outflow stops because diffusion can no longer supply the loss of hydrogen from advection. Last, we find that subsequent mass loss is diffusion-limited, yielding significantly lower mass loss rates than free advection, thus allowing planets to sustain hydrogen-rich atmospheres for significantly longer timescales.

\section*{Acknowledgements}

This work was sponsored by the US National Science Foundation EAR-2224727 and the US National Aeronautics and Space Administration under Cooperative Agreement {No.\,80NSSC19M0069} issued through the Science Mission Directorate. This work was also supported in part by the facilities and staff of the Yale University Faculty of Arts and Sciences High Performance Computing Center. We thank the anonymous reviewer, whose comments helped us to improve the clarity of our manuscript.

\bibliographystyle{agsm}
\bibliography{bibliography.bib}

\end{document}